\documentclass[12pt,a4paper]{article}
\usepackage{graphicx}
\usepackage{cite}
\usepackage{amssymb}
\title{Spatiotemporal perspective on the decay of turbulence in
wall-bounded flows}
\author{Paul Manneville\\
Laboratoire d'Hydrodynamique, Ecole Polytechnique\\
91128 Palaiseau, France\\
{\tt paul.manneville@polytechnique.edu}}

\date{Phys. Rev. E {\bf79} (2009) 025301 [R]  \&  039904 [E] }

\begin{document}
\sloppy

\maketitle
\begin{abstract}
Using a reduced model focusing on the in-plane dependence of plane Couette flow, it is shown that the {\it turbulent~$\to$~laminar\/} relaxation process can be understood as a nucleation problem similar to that occurring at a thermodynamic first-order phase transition.
The approach, apt to deal with the large extension of the system considered, challenges the current interpretation in terms of chaotic transients typical of temporal chaos.
The study of the distribution of the sizes of laminar domains embedded in turbulent flow proves that an abrupt transition from sustained spatiotemporal chaos to laminar flow can take place at some given value of the Reynolds number $R_{\rm low}$, whether or not the local chaos lifetime, as envisioned within low-dimensional dynamical systems theory, diverges at finite $R$ beyond  $R_{\rm low}$.\\[2ex]
\end{abstract}

The transition to turbulence in flows lacking linear instability modes, such as Poiseuille pipe flow driven by a pressure gradient along a circular tube (Ppf) and plane Couette flow driven by two plates moving parallel to each other in opposite directions (pCf), is particularly delicate to understand owing to its abrupt character, without the usual cascade seen in the {\it globally super-critical\/} case, as for e.g. convection.
A recent general presentation of the issues is given in \cite{MK05}.
These {\it globally sub-critical\/} flows become turbulent through the nucleation and growth or decay of turbulent domains called puffs (Ppf) or spots (pCf), see, e.g., \cite{Wetal08} and \cite{DHB92} for Ppf and pCf, respectively.
Most of the work on the transition problem has dealt with special nonlinear solutions (exact coherent structures \cite[b]{Wa97}) to the Navier--Stokes equations and their dynamics in the phase space spanned by them.
Such solutions, obtained within the so-called Minimal Flow Unit (MFU) assumption \cite{JM01}, have been found at moderate values of the Reynolds number $R$ in Ppf, \cite{FE03}, pCf \cite{Nl,CB97,Wa97,Wa03}, and also in plane Poiseuille flow \cite{IT01,Wa03}.
(In pCf, $R$ is defined as $Uh/\nu$ where $U$ is the speed of the plates driving the flow, $2h$ the gap petween the plates, and $\nu$ the kinematic viscosity of the fluid.)
These solutions are all unstable and, together with their stable and unstable manifolds, they form the skeleton of the turbulent flow at a local scale.
The reason why turbulence can be sustained only at much higher values of $R$ (about a factor of two to three higher) is however not clear \cite{Wetal08}.

In practice, the existence of these nontrivial solutions has mainly served to explain the exponential behavior of the distribution of lifetimes of transient turbulence in the low-$R$ part of the transitional regime in terms of unstable periodic orbits, homoclinic tangles and chaotic repellers \cite{Eetal07}.
Unfortunately, this does not bring a definitive answer to the controversial issue of the existence of a threshold $R_{\rm low}$ above which turbulence could be sustained, see \cite{Wetal08,PM06,Hetal06} for Ppf and \cite{Betal98,Hetal06} for pCf.
The reason is that the approach is zero-dimensional in essence (dynamics is condensed onto a small set of nonlinear modes with well defined spatial structure), which would be appropriate if confinement at the size of the MFU were really effective \cite[\S2.3]{Ma05}.
In fact, the transitional flows considered are open and, ideally, unbounded in one spatial direction (Ppf) or two (pCf), so that a genuinely {\it spatiotemporal\/} dynamics is expected.

A different angle of attack was proposed long ago by Pomeau \cite{Po86} (also \cite[Chap.~5]{BPV98}) who indeed stressed the importance of the spatial extension of systems experiencing subcritical bifurcations.
The coexistence of states in {\it phase space\/}, typical of subcriticality, translates into coexistence of states in {\it physical space\/}, with walls separating homogeneous domains of each kind.
He also pointed out that, when one of the competing states was chaotic and transient, one could expect a stochastic wall propagation similar to {\it directed percolation\/} \cite{Hi00}, another contamination process, with the obvious identification: locally {\it turbulent\/} $\to$ {\it active} (alive), locally {\it laminar\/} $\to$ {\it absorbing} (dead).
According to this scenario, called {\it spatiotemporal intermittency\/} (STI) by Kaneko \cite{Ka85}, transient chaotic local states of a distributed system may evolve into a sustained turbulent global regime due to spatial coupling.
STI has been much studied within the framework of critical phenomena in statistical physics, especially in view of its universality \cite{CM95}.
Coupled map lattices in different spatial dimensions were used to test STI's properties \cite{Betal01,Gr06}.
In two dimensions, such an abstract minimal model was found to display a discontinuous first-order like transition as the control parameter was varied \cite{CM88}, which was subsequently used for comparison with transitional pCf \cite[b]{Betal98}.
Unfortunately, it was too far removed from hydrodynamics to be really relevant, which prompted the derivation of a more realistic model directly from the Navier--Stokes equations, to be used here for studying statistical properties of the turbulent~$\to$~laminar transition in domains of very large lateral extension. 

A system of partial differential equations was derived by means of standard Galerkin expansion/projection of the Navier--Stokes equations, using a polynomial basis appropriate to no-slip boundary conditions; see \cite{LM07} for a full description and details on its numerical implementation.
This system extends Waleffe's model \cite{Wa97} by unfreezing the in-plane spatial dependence of the velocity field while keeping a few wall-normal modes.
When restricted to lowest significant order, as is done for results discussed here, the model involves three fields.
The first one describes {\it streaks\/} that are velocity perturbations pushing and pulling vortical structures.
The two others account for the rest of the flow, especially the {\it streamwise vortices\/} which, together with the streaks, are believed to form the main ingredients of the turbulence self-sustaining process \cite{Wa97,MK05}.
The model displays the most essential properties of the Navier--Stokes equations including lift-up, energy-preserving advection, and dissipation; it is also readily shown
that the base flow remains linearly stable for all Reynolds numbers.

When the streamwise length $L_x$ and spanwise width $L_z$ of the cell are large, simulations closely mimic the behavior of pCf in extended geometry.
A direct transition toward a homogenous turbulent state is observed at sufficiently large $R$, characterized by a turbulent energy density independent of $L_x,L_z\gtrsim16$ (lengths given in units of the half gap $h$) and by a uniform streamwise velocity correction giving its typical S shape to the turbulent mean flow \cite[a]{LM07}.
This essential correction is also present inside and around developing spots \cite[b]{LM07} as observed in the experiments, see also \cite[b]{BT05}.
However, due to the model's limited cross-stream resolution which underestimates energy transfer to, and dissipation in, small-scale wall-normal fluctuations, the transitional regime
observed in the model takes place at $R\sim170$, compared to $R\sim325$ in laboratory experiments. 

In the following, the discussion is limited to statistical properties of the model at the {\it turbulent~$\to$~laminar\/} transition as obtained from numerical simulations, using a conventional Fourier, fully de-aliased, pseudo-spectral code second order in time, appropriate for periodic in-plane boundary conditions.
Various spatiotemporal resolutions were used with similar results provided that numerical stability was guaranteed.
(A subcritical bifurcation toward a high-energy numerical mode was observed when the spatial step was too large.) 
Domains of various aspect ratios ranging from moderate to large, and then very large, have been considered.
All of our domains are much larger than the size of most active small scale structures in the system, which is $\approx6\times3$, the MFU.
The size of the experiments \cite{Betal98, Petal02} to which we compare our results ranged from $380\times70$ to $770\times340$. 

Moderate aspect ratios, $32\times32$, were first considered in \cite[a]{LM07}, showing turbulent transients with exponentially distributed lifetime probabilities.
The characteristic times of these distributions were observed to increase exponentially as in some Ppf experiments \cite{Hetal06} but seemingly faster for $R$ approaching $175$, suggesting a possible divergence around this value, i.e., the existence of a real threshold $R_{\rm low}\approx175$.
There were indications that this apparent cross-over could be due to spatiotemporal effects.

\begin{figure}
\begin{center}
\includegraphics[height=0.35\textwidth,clip]%
{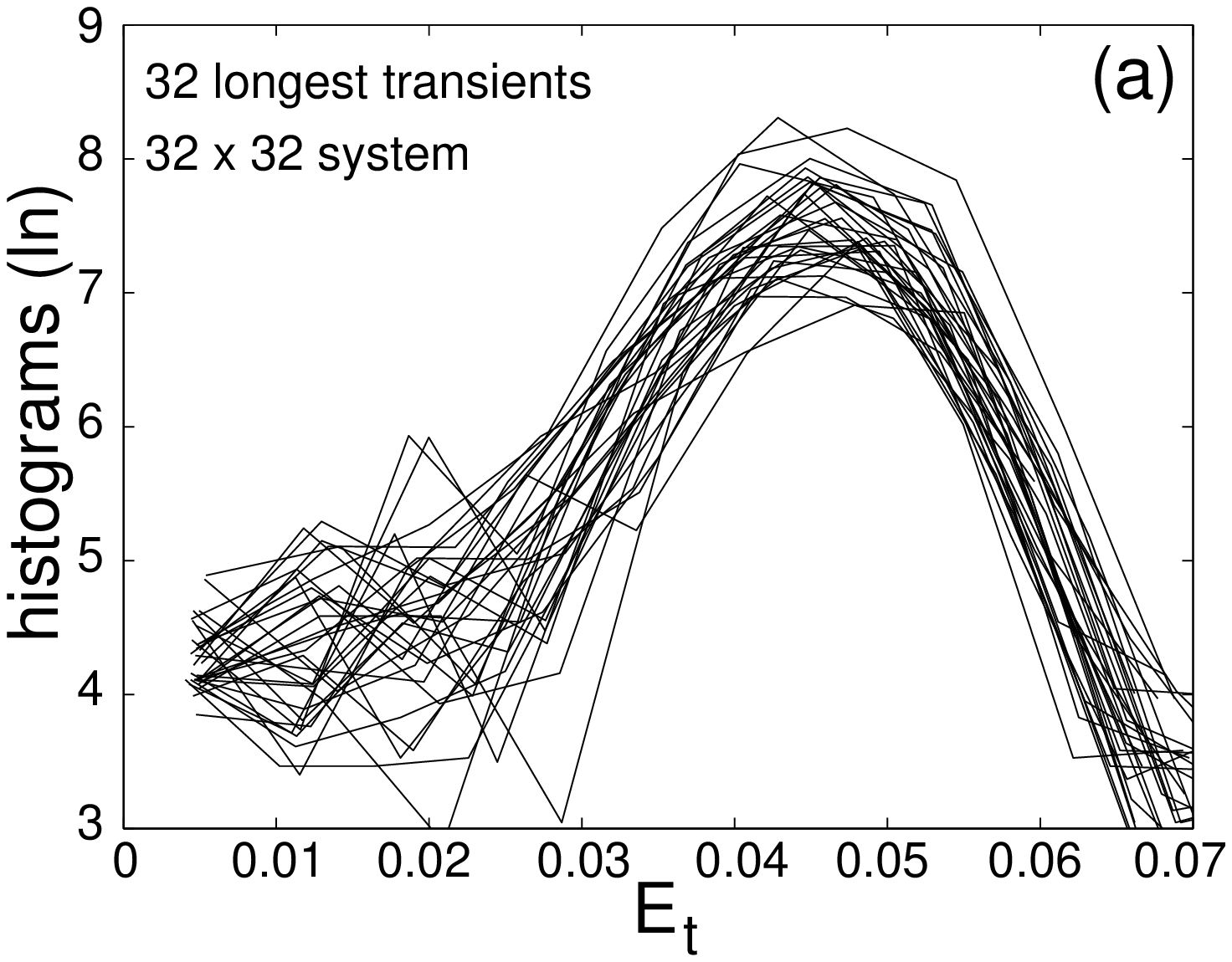}
\includegraphics[height=0.35\textwidth,clip]%
{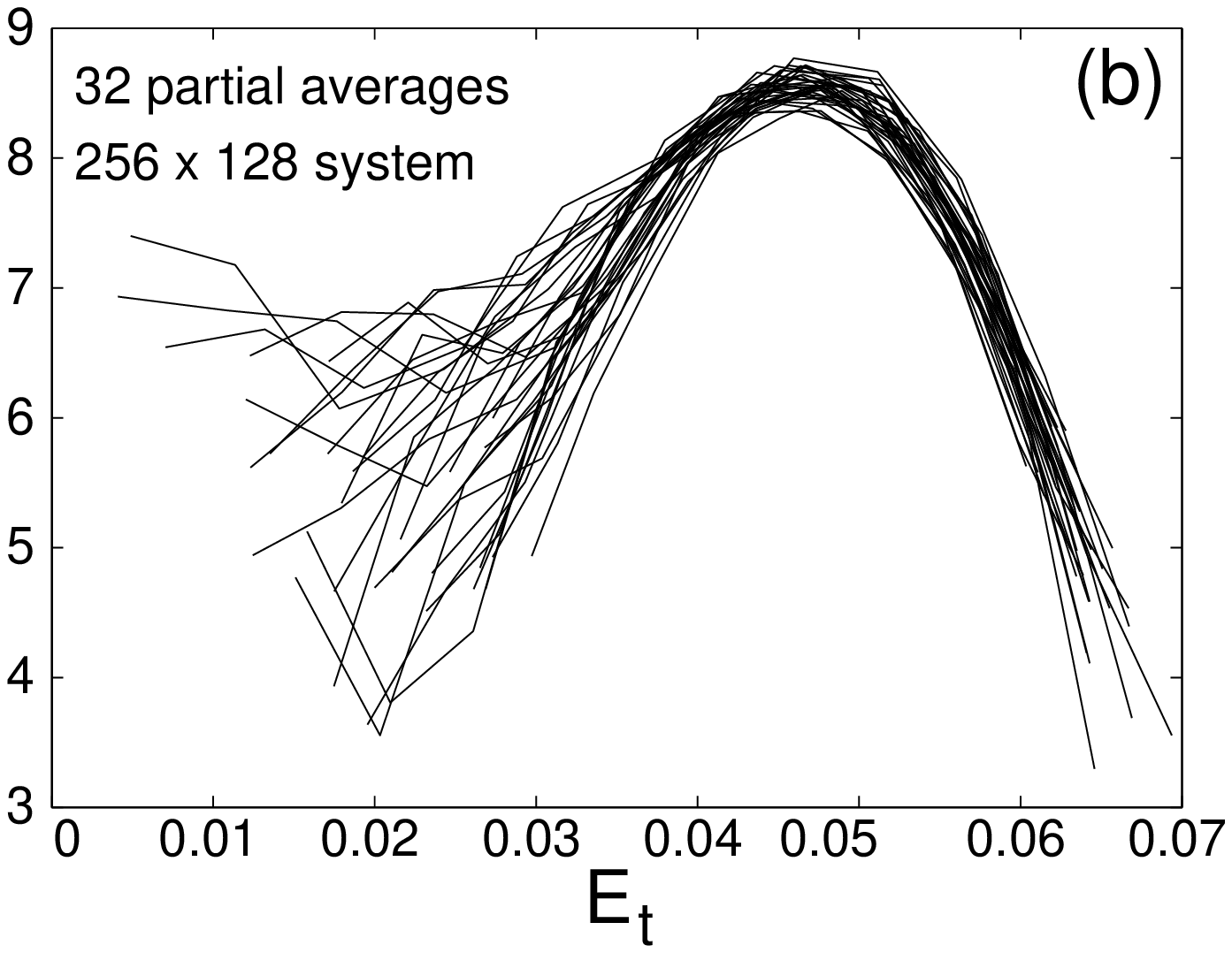}
\end{center}
\caption{\label{f1} Histograms of the values of $E_{\rm t}(t)$ during 32 transients in the $32\times32$ system (a) and for the 32 sub-domains covering the $256\times128$ system (b).}
\end{figure}

A larger domain, $256\times128$ was next considered \cite{Ma08}. It was first observed that the turbulent state could be maintained without difficulty down to $R\approx170$ over time intervals at least of the order of $3\times10^4$ advective time units, and that the average total perturbation energy per unit surface $E_{\rm t}$ fluctuated wildly \cite[Fig.~7, left]{Ma08},
with {\it highs\/} corresponding to a homogeneous distribution of small laminar patches immersed in turbulent flows and {\it lows\/} to a much more inhomogeneous situation with a few wide laminar domains surrounded by the same fine-grained mixture as for highs.
Fluctuations of $E_{\rm t}$ could then be attributed to the growth and decay of wide laminar domains.
In \cite[a]{LM07} it was shown that the turbulent state maintained itself in the $32\times32$ system as long as $E_{\rm t}(t)$ stayed above $\approx0.02$ (in units of $U^2$ per unit surface), while it inevitably decayed below.
Histograms of $E_{\rm t}(t)$ were constructed for 32 transients observed for $R=170$, Fig.~\ref{f1}(a), and compared to those of 32 time series of $E_{\rm t}$ obtained by averaging over 32 sub-domains of size $32\times32$ covering the $(256=8\times32)\times(128=4\times32)$ system at the same $R$, Fig.~\ref{f1}(b).
The histograms of the transient time series are less well defined than those of the sub-domains time series because the transients were always much shorter than the experiment in the $256\times128$ system.
All show similar humped shapes and identical behavior beyond their maxima but at, and below $E_{\rm t}\approx0.02$, an accumulation of values can be observed for the sub-domain time series whereas the histograms drop to zero for the transients.
Since the laminar patches observed in the $256\times128$ system were much wider than $32\times32$, the interpretation is obvious: a wide domain of size larger than $32\times32$ could become temporarily laminar ($E_{\rm t}\lesssim 0.02$) but was bound to return to the turbulent regime by {\it contamination\/} from its neighborhood while the small system could not recover (the transient ended), hence the link to directed percolation \cite{Po86}.

The first-order features of the transition have been examined by performing simulations in an even wider domain $\mathcal D=1536\times1536$, made possible thanks to the two-dimensional character of the model.
A slight shift of the global threshold, from $R_{\rm low}\approx170$ for $256\times128$ to $\approx171$, was the price to be paid for the lowering of the spatial resolution by a factor of three ($\delta x=\delta z=0.75$,  i.e., $12\times6$ collocation points per MFU) and of the temporal resolution by a factor of 20 ($\delta t=0.2$, i.e. five computations per turn-over time).
This permitted data accumulation over long durations, typically of the order of $10^5$ advective time units.

\begin{figure}
\begin{center}
\includegraphics[width=0.75\textwidth]{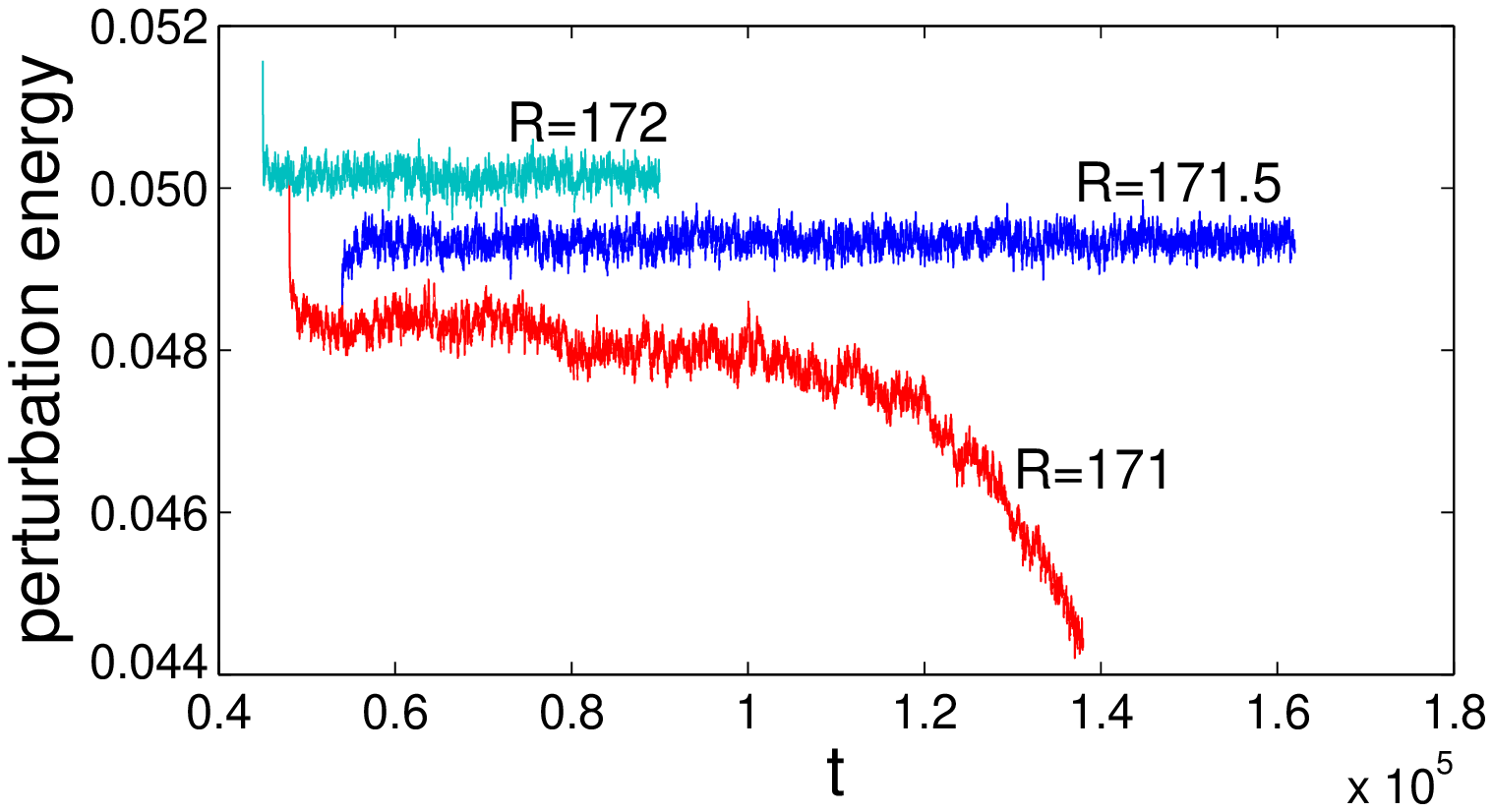}
\bigskip\bigskip

\includegraphics[width=0.4\textwidth]{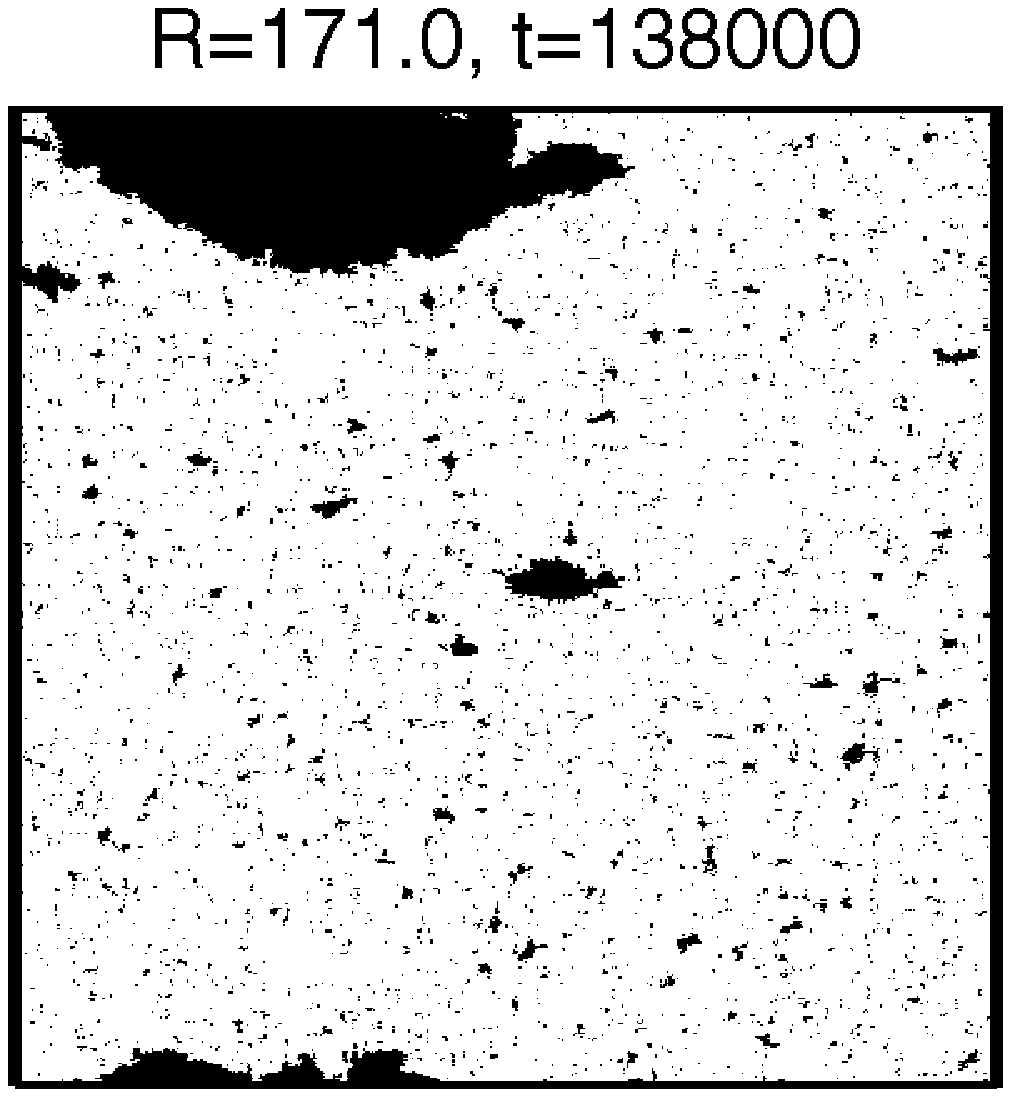}\hspace{1em}\includegraphics[width=0.4\textwidth]{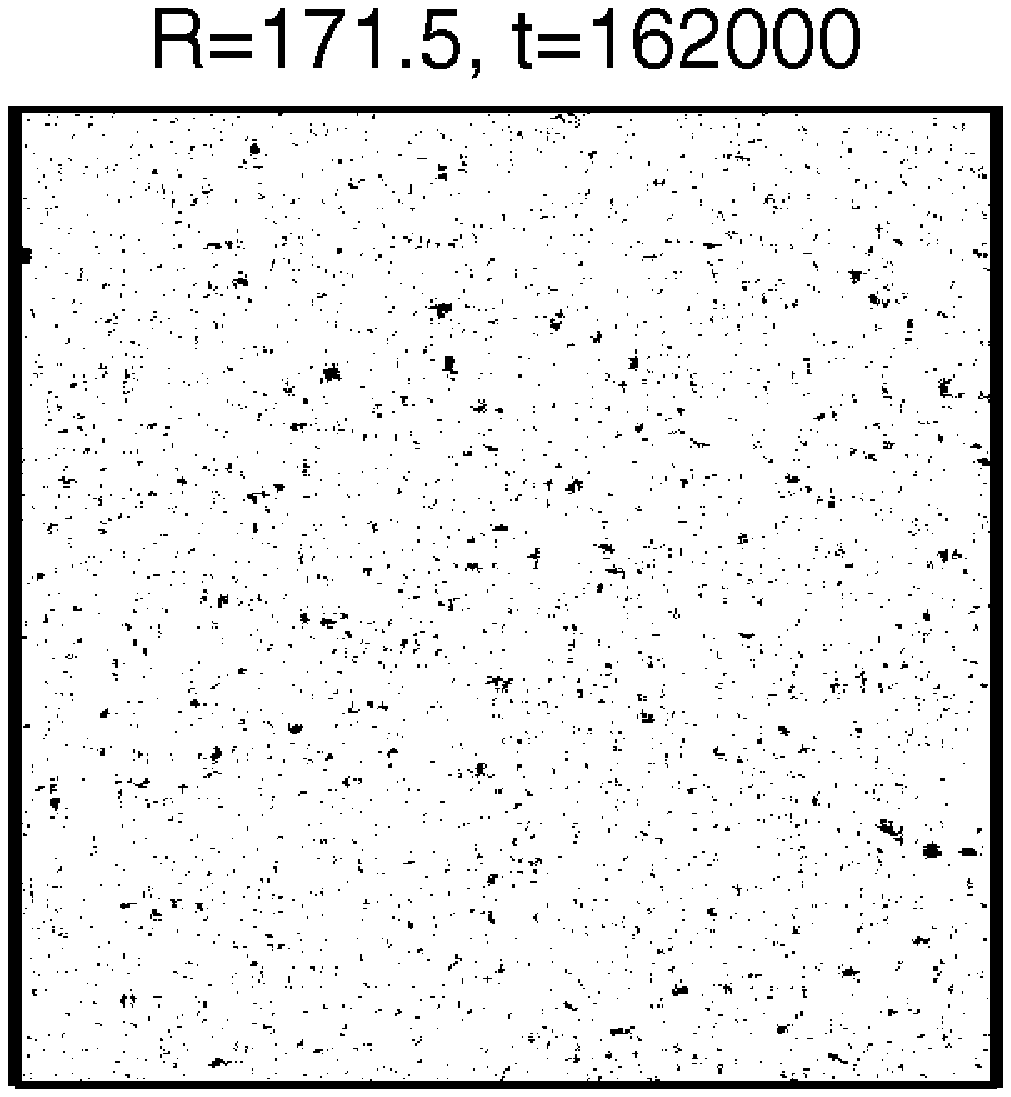}
\end{center}
\caption{\label{f2} Top: time series of $ E_{\rm t}$ for the extra-wide domain $\mathcal D=1536\times1536$. Bottom: two b/w snap-shots of the solution for $R=171$ (left) and $R=171.5$ (right). Thresholding performed according to the rule given in the text.}
\end{figure}

Figure~\ref{f2} (top) displays time series of the energy per unit surface $ E_{\rm t}$ contained in the perturbation and averaged over the whole domain $\mathcal D$ for three experiments close to the global stability threshold.
Experiments were performed by decreasing $R$ from higher values to $R=172$, then to $R=171$ where a large laminar domain invaded the system, Fig.~\ref{f2} lower left, and finally increasing $R$ back to $R=171.5$, for which all wide laminar domains receded leaving a whole distribution of small laminar patches, Fig.~\ref{f2} lower right.
The turbulent state then remained over a long period of time without any sign of decay.

\begin{figure}
\includegraphics[height=0.38\textwidth,clip]{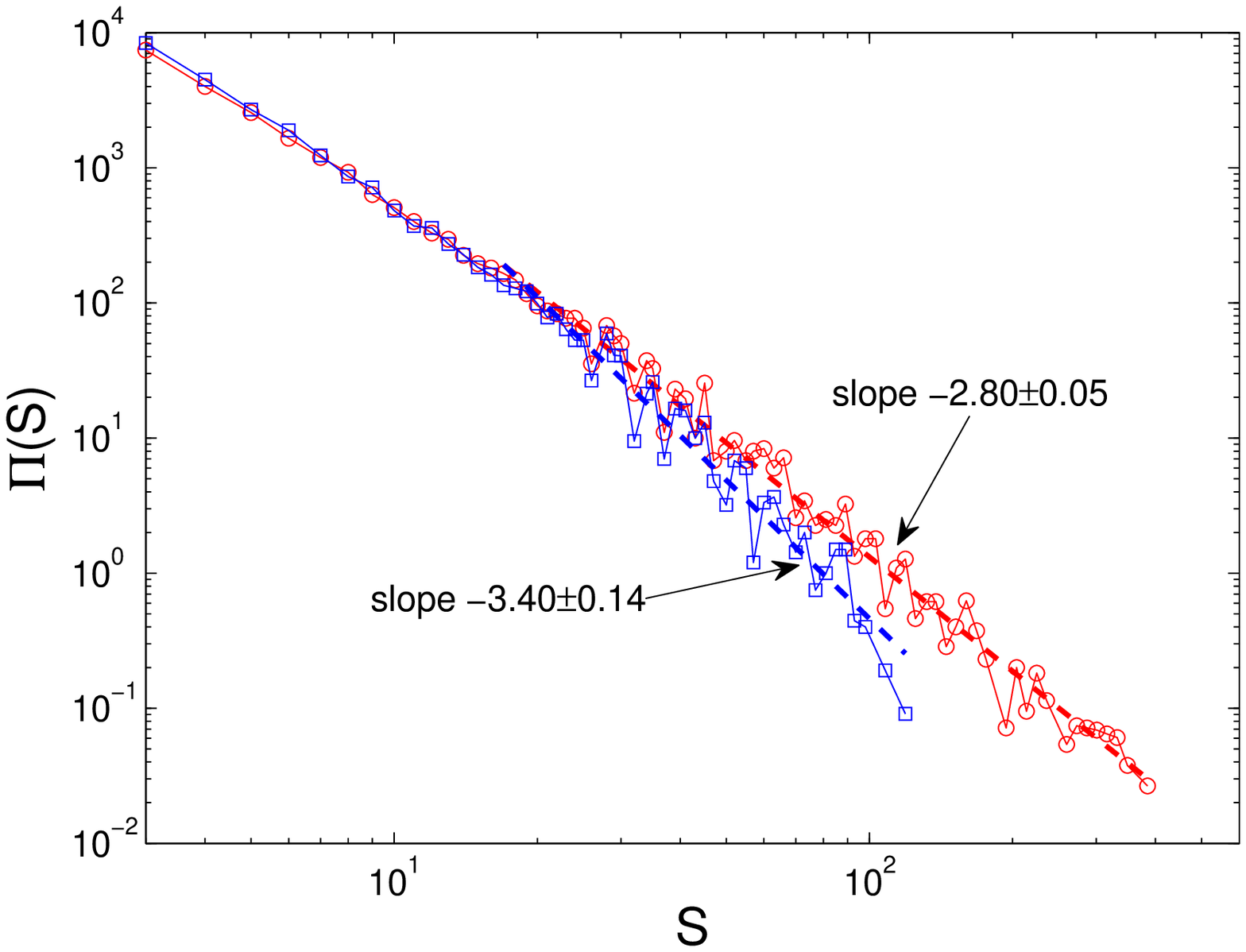}\hfill
\includegraphics[height=0.38\textwidth,clip]{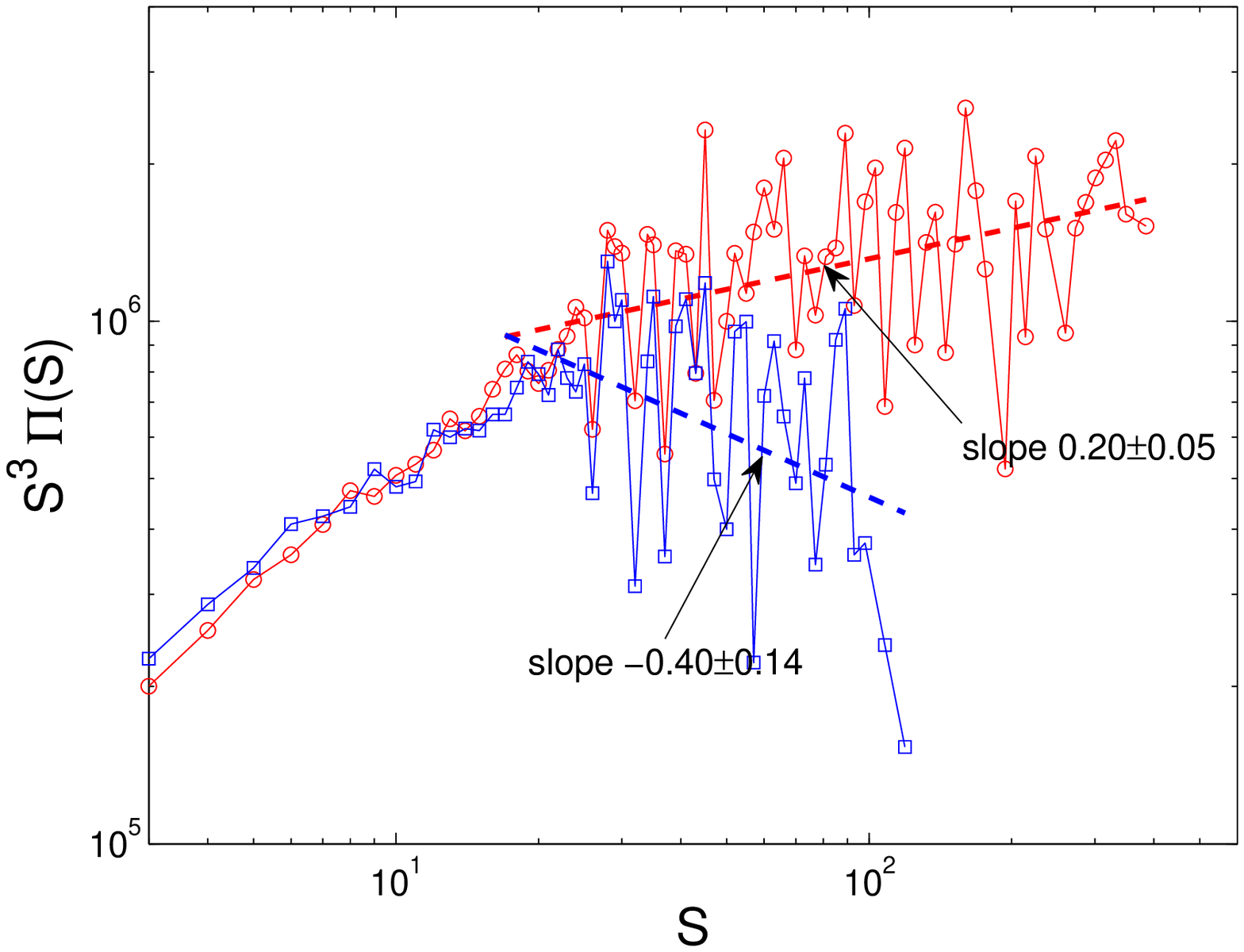}
\caption{\label{f3}  Left:
Probability distribution of cluster areas $\Pi(S)$ as a function of the surface $S$ (numbers of laminar pixels). Right: $S^3\Pi(S)$ as a function of $S$. $R=171.5$: red/squares. $R=171$:
blue/circles.}
\end{figure}

The statistical analysis of the flow patterns next proceeded with the detection of laminar domains.
The local perturbation energy  was coarse-grained over pixels of size $3\times3$ (i.e. $1/2$ MFU).
A pixel was declared {\it laminar\/} when this coarse grained quantity was below $0.01$ and {\it turbulent\/} otherwise.
The coarse graining procedure avoided the spurious detection of stagnation points in the velocity field.
A Hoshen--Kopelman algorithm \cite{HK76} as implemented by Domany \cite{httpDo} was used to identify the laminar clusters which were sorted according to their area measured in terms of pixels.
Ranking clusters by decreasing size immediately gives cumulated distributions.
The probability distributions shown in Figure~\ref{f3} were obtained by differentiating the cumulated distributions.
They were then slightly smoothed by resampling on bins of widths geometrically increasing with a factor 1.05.
The pixel size, the ``on/off'' threshold and the smoothing factor resulted from a tedious trial-and-error procedure guided by the contradictory requirements of noise-level reduction and significance of the results.
On the one hand, in contrast to the noise level, the slopes were not found sensitive to the value of the pixel-size/threshold combinations. On the other hand, the bin resizing factor was kept as close to one as possible in order to preserve the average value of the slopes visible in the raw data.

Probability distribution of laminar cluster areas for $R=171.5$ and $R=171$ are displayed in Figure~\ref{f3}.
The distribution for $R=172$ resembles that for $R=171.5$; extending a little less to large sizes, it is not presented for the sake of clarity.
At higher values of $R$ very few large laminar domains were detected and the distributions decayed faster, making quantitative estimates unreliable.
All these distributions were obtained by scanning a series of b/w snapshots such as those in Figure~\ref{f2} bottom part.
Since they were taken every $\Delta t=3\times10^3$, the snapshots were sufficiently spaced in time to be statistically independent, whereas the large invading cluster correlating successive snapshots for $R=171$ was systematically discarded from the statistics.
All the distributions collapsed for laminar domains with areas below ten pixels ($\sim 5$ MFUs).
Tails with power law behavior and $R$-dependent slopes were observed for the largest sizes.

While the first momentum of the distribution of laminar domain sizes, $\langle S \rangle = \int_{S_{\rm min}}^\infty S\,\Pi(S)\, {\rm d}S$, remains defined, its second moment, $\langle S^2 \rangle = \int_{S_{\rm min}}^\infty S^2\,\Pi(S)\,{\rm d}S$, may diverge depending on whether or not the exponents $\alpha$ of the tails of the distributions are smaller or larger than 3.
From the nucleation viewpoint, the divergence of the variance is in fact more meaningful than the divergence of the mean erroneously claimed in the version published initially \cite{Ma09}. 
Distributions of laminar domain sizes can be defined on time scales much longer than those of turbulent pattern renewal.
If a laminar germ grows without bounds when it is larger than some critical size and recedes otherwise, then, as long as the variance remains somewhat smaller than the critical germ size,
the probability of occurrence of a critical germ is negligible and turbulence remains sustained ($R=171.5$, $\alpha\simeq3.4$).
When the variance diverge, a critical germ, whatever its size, will appear with finite probability and next will invade the system:
turbulence is bound to decay for $R=171$ ($\alpha\simeq2.8$). The change of behavior is expected for $171.5>R>171$ in our model and is conjectured to take place at $R\sim325$ in Couette flow as reported by Bottin {\it et al.} \cite{Betal98}, despite the reinterpretation of experimental data by Hof {\it et al.} \cite{Eetal07}. 

It might be argued that, despite its appealing properties, the model is not yet Navier--Stokes. In particular, due to its lack of cross-stream resolution, it cannot give information about the upper part of the transitional regime where oblique turbulent bands are observed \cite{Petal02,BT05}.
However it points toward a scenario that is likely to work in subcritical wall flows such as Ppf (quasi-1D), pCf (quasi-2D), or even the plane Poiseuille flow and the boundary layer flows (also quasi-2D) that all display regimes of laminar/turbulent coexistence at intermediate Reynolds numbers.
STI is the process by which {\it transient local\/} chaos is converted into {\it sustained global\/} spatiotemporal chaos.
Two ingredients are necessary: a controllable decay rate of local states
and a coupling with neighbors in physical space.
Now, assume that, within the MFU assumption, the well-documented transient chaotic behavior has its lifetime primarily tuned by the Reynolds number.
When they appear, nontrivial solutions have a short lifetime which leaves no time for contamination to work efficiently and so the flows relax to laminar.
When $R$ increases, the local deterministic dynamics in each MFU becomes more complicated and, as a result, the lifetime of transients increases.
Due to the coupling, a genuine transition to sustained spatiotemporal chaos (weak turbulence) takes place at some given value $R_{\rm low}$, {\it whether or not the local lifetime, as obtained from the low-dimensional dynamical systems theory, diverges for $R_{\rm low} < R < \infty$}.
This transition would be within the class defined by STI and could be second-order (continuous) or first-order (discontinuous).
Experiments tell us that, in the cases studied so far, it is first-order and thus deprived  of any universality (correlation lengths remain finite at threshold).
Abstract models considered up to now have been designed on lattices with local couplings
and display both kinds of transitions in both one or two dimensions \cite{CM88,Betal01,CM95,Gr06}.
The problem with this picture ---solved by our (semi-)realistic Navier-Sokes modeling of pCf--- is that couplings of hydrodynamic origin may have non-local effects linked to pressure, which makes size an important issue.
This should urge us to develop modeling further, not only in quasi-2D as was done here for pCf, but also in the quasi-1D case of Ppf by pushing further the simplification effort made in \cite{WKxx} in order to capture the spatiotemporal essence of the transition \cite{Ba11}.

\paragraph{Acknowledgments:} 
Fruitful interactions with M.~Lagha,
C.~Cossu, J.M.~Chomaz, H.~Chat\'e,
O.~Dauchot, F.~Daviaud, B.~Eckhardt,
L.~Tuckerman, R.R.~Kerswell,  T.~Mullin, F.~Waleffe,  and many others,
are deeply acknowledged.
Numerical simulations were performed thanks to CPU allocations
of IDRIS (Orsay) under projects \#61462, \#72138.
This article was prepared during a stay at the Isaac Newton
Institute (Cambridge) and those who took part in the workshop
{\it Wall bounded shear flow: transition and turbulence\/} should be thanked for discussions, criticisms, and encouragements.
The author is indebted to H.~Chat\'e for pointing out the faulty data treatment in the version published initially, and for crucially contributing to the revised interpretation.


\begin{thebibliography}{38}

\bibitem{MK05}
T. Mullin and R. Kerswell, eds.,
\emph{IUTAM Symposium on Laminar--Turbulent Transition and
Finite Amplitude Solutions} (Springer, Dordrecht, 2005).

\bibitem{Wetal08}
A. Willis, J. Peixhino, R. Kerswell, and T. Mullin, Phil. Trans. R. Soc.  A \textbf{366}, 2671 (2008).

\bibitem{DHB92}
F. Daviaud, J. Hegseth and P. Berg\'e, Phys. Rev. Lett. \textbf{69}, 2511 (1992).
O. Dauchot and F. Daviaud, Phys. Fluids A \textbf{7}, 335 (1995).

\bibitem{Wa97}
F. Waleffe, a) Phys. Fluids \textbf{9}, 883 (1997);
b) Phys. Rev. Lett. \textbf{81}, 4140 (1998).

\bibitem{JM01}
J. Jimenez and P. Moin, J. Fluid Mech. \textbf{225}, 213 (1991).

\bibitem{FE03}
H. Faisst and B. Eckhardt, Phys. Rev. Lett. \textbf{91}, 224502 (2003).
H. Wedin and R. Kerswell, J. Fluid Mech. \textbf{508}, 333 (2004).
C. Pringle, Y. Duguet and R. Kerswell,  Phi. Trans. R. Soc. {\bf 367}, 547 (2009).

\bibitem{Nl}
M. Nagata, J. Fluid Mech. \textbf{217}, 519 (1990).
G. Kawahara and S. Kida, J. Fluid Mech. \textbf{449}, 291 (2001).
D. Viswanath, J. Fluid Mech. \textbf{580}, 339 (2007).

\bibitem{Wa03}
F.~Waleffe, Phys. Fluids \textbf{15}, 1517 (2003).

\bibitem{CB97}
R. Clever and F. Busse, J. Fluid Mech. \textbf{344}, 137 (1997).

\bibitem{IT01}
T. Itano and S. Toh, J. Phys. Soc. Japan \textbf{70}, 703 (2001).

\bibitem{Eetal07}
B. Eckhardt, T. Schneider, B. Hof, and J. Westerweel, Ann. Rev. Fluid Mech. \textbf{39}, 447 (2007).

\bibitem{Hetal06}
B. Hof, J. Westerweel, T. Schneider, and B. Eckhardt, Nature \textbf{443}, 59
(2006).

\bibitem{PM06}
J. Peixinho and T. Mullin, Phys. Rev. Lett. \textbf{96}, 094501 (2006).
A. Willis and R. Kerswell, Phys. Rev. Lett. \textbf{98}, 014501 (2007).

\bibitem{Betal98}
a) S. Bottin, F. Daviaud, P. Manneville, and O. Dauchot, Europhys. Lett. \textbf{43}, 171 (1998);
b) S. Bottin and H. Chat\'e, Eur. Phys. J. B \textbf{6}, 143 (1998).

\bibitem{Ma05}
P. Manneville, in [1],  pp. {1--33}.

\bibitem{Po86}
Y. Pomeau, Physica D \textbf{23}, 3 (1986).

\bibitem{BPV98}
P. Berg\'e, Y. Pomeau, and C. Vidal,
\emph{L'espace chaotique} (Hermann, Paris,1998).

\bibitem{Hi00}
H. Hinrichsen, Advances in Physics \textbf{49}, 815 (2000).

\bibitem{Ka85}
K. Kaneko, Prog. Theor. Phys. \textbf{74}, 1033 (1985).

\bibitem{CM95}
H. Chat\'e and P. Manneville, in
\emph{Turbulence: a Tentative Dictionary}, P. Tabeling and O.~Cardoso, eds.
(Plenum Press, New York, 1995), pp. 111--116.

\bibitem{Betal01}
T. Bohr, M. van Hecke, R. Mikkelsen, and M. Ipsen, Phys. Rev. Lett. \textbf{86}, 5482 (2001).

\bibitem{Gr06}
P. Grassberger, J. Stat. Mech. \textbf{6}, P01004 (2006).

\bibitem{CM88}
H. Chat\'e and P. Manneville, Europhys. Lett. \textbf{6}, 591 (1988).

\bibitem{LM07}
M. Lagha and P. Manneville, a) Eur. Phys. J B \textbf{58}, 433 (2007);
b) Phys. Fluids \textbf{19}, 094105 (2007).

\bibitem{BT05}
D. Barkley and L. Tuckerman,
a) Phys. Rev. Lett. \textbf{94}, 014502 (2005);
b) J. Fluid Mech. \textbf{576}, 109 (2007).

\bibitem{Petal02}
A. Prigent, G. Gr\'egoire, H. Chat\'e, O. Dauchot, and W. van Saarloos, Phys. Rev. Lett. \textbf{89}, 014501 (2002).

\bibitem{Ma08}
P. Manneville, Pramana --- Journal of Physics
\textbf{70}, 1009 (2008).

\bibitem{HK76}
J. Hoshen and R. Kopelman, Phys. Rev. B \textbf{14}, 3438 (1976).

\bibitem{httpDo}
E. Domany, (1999), http://www.weizmann.ac.il/home/fedomany/RG99/coarsening.m.

\bibitem{WKxx}
A. Willis and R. Kerswell, J. Fluid Mech. {\bf 619}, 213 (2009).

\bibitem{Ma09}
P. Manneville, Phys. Rev. E {\bf79}, 025301 [R] (2009).

\bibitem{Ba11} Note (2011/02/28): This program has now been achieved in a large part by D.~Barkley, Simplifying the complexity of pipe flow arXiv:1101.4125v1 [physics.flu-dyn], and private communication.

\end{thebibliography}
\end{document}